\documentclass[11pt]{llncs}
\usepackage[breaklinks=true,letterpaper=true,colorlinks,citecolor=black,bookmarks=false,]{hyperref}
\usepackage{amsmath,amssymb}
\usepackage{enumitem}
\usepackage[sort&compress,numbers]{natbib}
\usepackage[normalem]{ulem}

\usepackage{times}
\usepackage{graphicx} 
\usepackage{subcaption}
\usepackage[noend,linesnumbered]{algorithm2e}

\usepackage[letterpaper, margin=1.2in]{geometry}

\title{DeepIso: A Deep Learning Model for Peptide Feature Detection}
\author{Fatema Tuz Zohora\inst{1} \and Ngoc Hieu Tran\inst{1} \and Xianglilan Zhang\inst{1} \and  Lei Xin\inst{2} \and \\Baozhen Shan\inst{2} \and Ming Li\inst{1}\thanks{Correspondence and 
requests for materials should be addressed to Ming Li (Email: mli@uwaterloo.ca)}}

\institute{David R. Cheriton School of Computer Science, University of Waterloo, Waterloo, Ontario, Canada 
\and Bioinformatics Solutions Inc., Waterloo, Ontario, Canada}

\begin{document}
\maketitle

\begin{abstract}
Liquid chromatography with tandem mass spectrometry (LC-MS/MS) based proteomics is a well-established research field with major applications such as identification of disease biomarkers, drug discovery, drug design and development. In proteomics, protein identification and quantification is a fundamental task, which is done by first enzymatically digesting it into peptides, and then analyzing peptides by LC-MS/MS instruments. The peptide feature detection and quantification from an LC-MS map is the first step in typical analysis workflows. In this paper we propose a novel deep learning based model, DeepIso, that uses Convolutional Neural Networks (CNNs) to scan an LC-MS map to detect peptide features and estimate their abundance. Existing tools are often designed with limited engineered features based on domain knowledge, and depend on  pretrained parameters which are hardly updated despite huge amount of new coming proteomic data. Our proposed model, on the other hand, is capable of learning multiple levels of representation of high dimensional data through its many layers of neurons and continuously evolving with newly acquired data. 
To evaluate our proposed model, we use an antibody dataset including a heavy and a light chain, each digested by Asp-N, Chymotrypsin, Trypsin, thus giving six LC-MS maps for the experiment. Our model achieves 93.21\% sensitivity with specificity of $99.44\%$ 
on this dataset. Our results demonstrate that novel deep learning tools are desirable to advance the state-of-the-art in protein identification and quantification.
\end{abstract}


\pagebreak
\section{Introduction}\label{sec:intro}
The outstanding performance of deep learning on object recognition opens a new frontier in the domain of bioinformatics. As a continuation of that, our research on solving peptide feature detection problem using Convolutional Neural Network (CNN) is the first attempt as per our knowledge. 
The use of CNN in image processing is pioneered by Yann LeCun et al.~\cite{lecun1998gradient} in 1998, for the hand written digit recognition. However, CNN became popular after the revolutionary breakthrough in 2012 ImageNet~\cite{krizhevsky2012imagenet} object recognition competition\footnote{https://www.technologyreview.com/s/530561/the-revolutionary-technique-that-quietly-changed-machine-vision-forever/}.
\\

On the other hand, proteomics based on liquid chromatography with tandem mass spectrometry (LC-MS/MS) is a well established technology for discovery of disease biomarkers, drug target identification, mode of action (MOA) studies and safety marker identification in drug research~\cite{aoshima2014simple}. Protein identification and quantification are fundamental tasks in proteomics and peptides are the building block of protein. Therefore, the typical analysis workflows of LC-MS/MS data include peptide feature detection and quantification from an LC-MS map, peptide identification from MS/MS spectra, and protein profiling~\cite{steen2004abc, zhang2012peaks, sturm2008openms}. The first step, peptide feature detection and quantification from an LC-MS map is our target problem. The LC-MS map of a protein sample is a 3D plot where the three dimensions are: mass-to-charge ($m/z$) or Da, retention time (RT), and intensity of peptide ions in that sample. Peptide feature is a multi-isotope pattern formed by different molecular isotopes, e.g. carbon-12 and carbon-13, of the same peptide. Detecting multi-isotope patterns in LC-MS map is a  challenging task due to the overlapping peptides, several charges of the same molecule, and intensity variation. Moreover, a single LC-MS map may have gigapixel dimension containing thousands to millions of  peptide features. However, CNN is found to be effective in similar pattern recognition problems, for example, in detecting cancer metastasis on gigapixel pathology images by Liu et al.~\cite{liu2017detecting}. Therefore, to address our target problem, we propose a new model DeepIso, based on CNN, that slides a window detector over the LC-MS map to spot multi-isotope patterns. The goal is to detect peptide features along with their charge states, and estimate their intensities. 
\\

Latest advanced types of LC-MS technologies generate huge amounts of analytical data with high scan speed, high accuracy and resolution, which is almost impossible to interpret manually. 
Existing methods to automate this data handling applies different heuristics and none of them relies on deep learning to find out the appropriate parameters automatically from the available LC-MS data. For example, in MaxQuant~\cite{cox2008maxquant}, peaks (component of a peptide feature) are detected by fitting a Gaussian peak shape, and then the peptide feature is found by employing a graph theoretical data structure. AB3D~\cite{aoshima2014simple} first roughly picks all local maxima whose intensity is larger than a given threshold  as candidate peaks from the entire LC-MS map. Then applies an iterative algorithm to process neighboring peaks of each candidate peak, to form peptide feature. Their recall varies from 0.35 to 0.85, and precision varies from 0.14 to 0.53 
based on different datasets. MSight~\cite{palagi2005msight} generates images from the raw MS data file for adapting the image-based peak detection. CentWave~\cite{tautenhahn2008highly} uses a pre-scan to first identify regions of interest composed of centroids and then the centroids are collapsed into a one-dimensional chromatogram, and wavelet-based curve fitting is performed to separate closely eluting peaks. Their F-score varies from 55\% to 85\% based on different dataset. TracMass~\cite{tengstrand2014tracmass} and Massifquant~\cite{conley2014massifquant} uses a 2D Kalman Filter (KF) to find peaks in highly complex samples. Massifquant's sensitivity varies from 75\% to 90\% and specificity varies from 80\% to 100\% based on different datasets. 
\\

In most of the existing algorithms, many parameters are set based on experience with empirical experiments, whose different settings may have a large impact on the outcomes.
In contrast to these existing works, our research aims at systematically training CNN using real dataset to automatically learn all characteristics of the data, without human intervention. Last but not least, even if the model makes wrong predictions, the correct results can be put back as new training data so that the model can learn from its own mistakes. We believe that such models shall have superior performance over existing techniques and shall become the method of choice in the near future.

\section{Method}\label{basics}
\begin{figure}[h!]
\centering
\includegraphics[scale = .4]{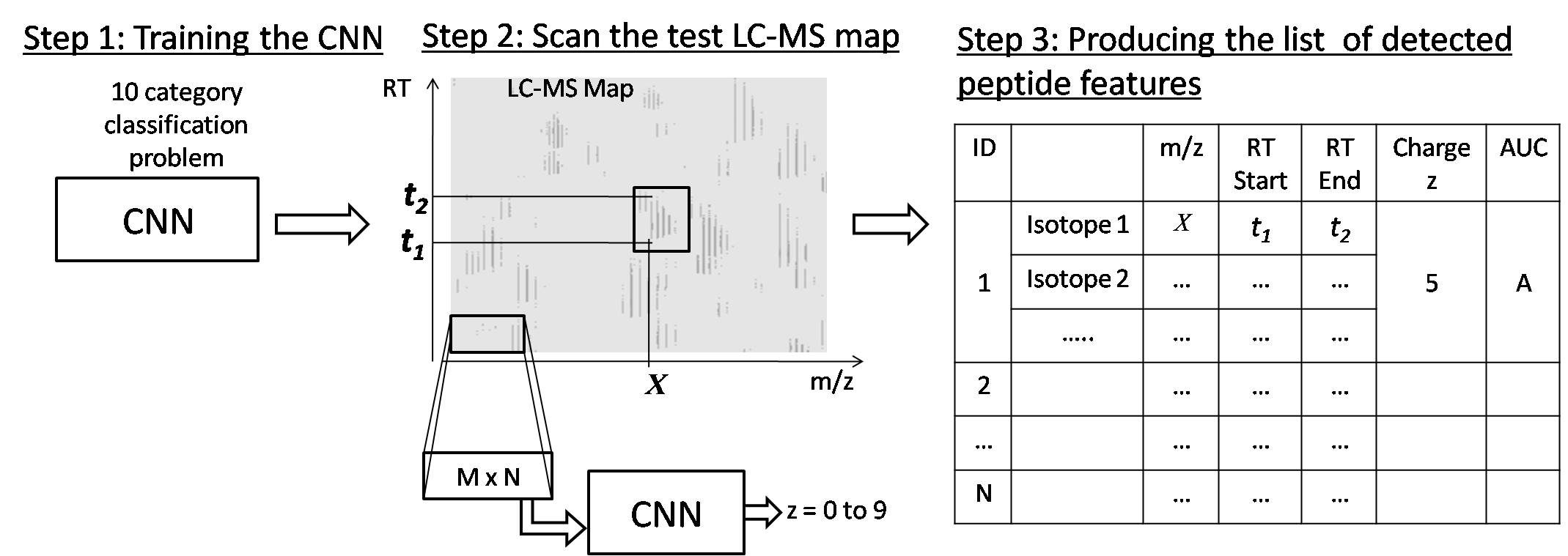}
\caption{Block diagram of our proposed method to detect peptide features from LC-MS map of protein sample}\label{alg}
\end{figure}

We explain our method using a block diagram as shown in Figure~\ref{alg}. The method can be divided into three steps. First step is to train the CNN, second step is to scan the test LC-MS map to detect the peptide features using the trained model in first step, and third step is to process those detections to produce a list of peptide features. First two steps involve deep learning and the third step applies heuristics. Testing phase involves Step 2 and Step 3. In the following sections we discuss each of the steps in detail.

\subsection{Step 1: Training The CNN}
The intuition of the first step is that, CNN takes a $[M\times N]$ dimension image (cut from training LC-MS maps) as input and outputs `Yes' or `No', based on whether it sees a peptide feature in the image or not. If the output is `Yes', then the CNN also needs to identify the charge $z$ of that feature. The $z$ can have values from 1 to 9. Therefore we use $z=0$ as an indication of `No' feature is seen in the input image. So this training step is basically a supervised 10 category classification problem using CNN, where each category is a charge $z=$ 0 to 9. In this step, CNN is supposed to learn following basic properties of peptide feature~\cite{cappadona2012current}, besides many other hidden characteristics from the training data.
\begin{enumerate}
\item In the LC-MS map, the isotopes in a peptide feature are equidistant along $m/z$ axis. For charge $z=$ 1 to 9, the isotopes are respectively 1.00 $m/z$, 0.5 $m/z$, 0.33 $m/z$, 0.25 $m/z$, 0.17 $m/z$, 0.14 $m/z$, 0.13 $m/z$, and 0.19 $m/z$ distance apart from each other.
\item The intensities of the isotopes form bell shape within their retention time (RT) range. 
\item The isotope having highest intensity in a peptide feature is called precursor ion and usually that is the first isotope in a feature. For instance, isotope 1 listed in the feature table shown in Figure~\ref{alg}. 
\item Peptide features often overlap with each other. 
\end{enumerate}

\subsection*{Training Data Generation}
We use the dataset WIgG1 of monoclonal antibody sequence including a light chain and a heavy chain~\cite{tran2016complete} to perform the experiment. It was generated from the LC-MS/MS analysis of the Intact mAb Mass Check Standard purchased from Waters. It is an intact mouse antibody purified by Protein-A with known molecular weights and amino acid sequences of both the light and heavy chains. Since each chain is digested by Asp-N, Chymotrypsin, and Trypsin, therefore we have in total six LC-MS maps for doing the experiment\footnote{The RAW files of the antibody dataset can be downloaded from the database MassIVE with accession number MSV000079801}. We produce a list of peptide features from each map using PEAKS Studio\footnote{http://www.bioinfor.com/peaks-studio/}, and consider that as our ground truth.
\\

We apply 6-fold cross validation on these six LC-MS maps. Each time we keep one map for testing and the remaining for the training. Each map holds about 20,000 peptide features. We consider resolution of 0.01 $m/z$ along the horizontal axis, and 0.01 minute along the vertical axis. Based on this resolution, each LC-MS map has dimension of around $[2,000 \times 100,000]$ pixels. We scale the pixel intensities in each map from 0 to 255. For clarification please refer to the LC-MS map shown in Step 2 in Figure~\ref{alg}. We consider the features having charge $z=$ 1 to 9 as positive samples and $z=0$ as negative samples (no feature present). We cut the features from the training LC-MS maps, considering a block/window size of $[M\times N]=[15 \times 211]$. In Appendix A1 we discuss why this block size is good enough to cover a feature so that CNN can take decision about the existence and charge of a feature.
\\

\begin{figure}
\begin{subfigure}{.5\textwidth}
  \centering
  \includegraphics[width=.85\linewidth]{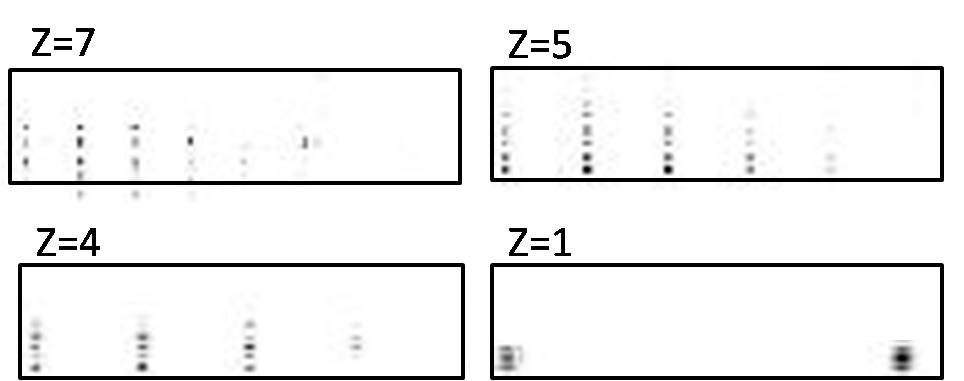}
  \caption{}
  \label{pozs}
\end{subfigure}%
\begin{subfigure}{.5\textwidth}
  \centering
  \includegraphics[width=.9\linewidth]{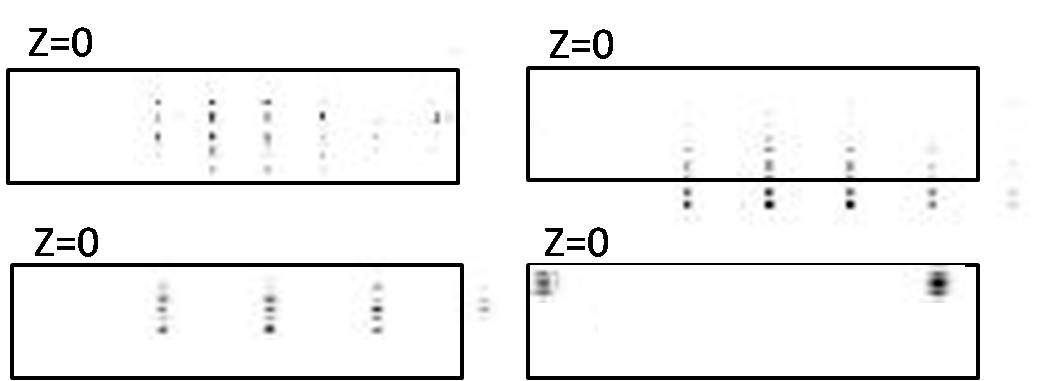}
  \caption{}
  \label{negs}
\end{subfigure}
\caption{Generation of training data: (a) Generate positive samples by placing a [15 x 211] window over the feature, such that, first  isotope of the feature is centered at [0,6] pixel of the window; (b) Generate negative samples by translating the window around the features, such that, NO feature starts within the $[0, 0]$ to $[0, 6]$ pixels of the window}
\label{fig:fig}
\end{figure}

We cut the positive features so that the feature is placed at pixel $[0,6]$ of the block as shown in Figure~\ref{pozs}, because some features have wider isotopes, for instance, the bottom right feature shown in the figure. The amount of features having charge $z=1,2,3,4$ is higher than other positive samples in the dataset. Therefore to make the dataset balanced, we add some synthetic data for charge 5 to 9. The procedure of synthetic data generation is explained in Appendix A4. To generate negative samples, we select some features and cut blocks from its surrounding region satisfying the condition that NO feature starts within the $[0, 0]$ to $[0, 6]$ pixels of the block as shown in Figure~\ref{negs}. 
\\

In this way we cut about 55,000 positive samples and about 90,000 negative samples for each fold, giving about 1.4 million features for training where the percentage of features having $z=$ 0 to 9 is about 62\%, 7\%, 10\%, 9\%, 4\%, 5\%, 2\%, 0.5\%, 0.2\%, 0.1\% respectively. We select 20\% of them for validation such that, validation dataset \emph{does not contain} synthetic positive samples.  The ratio of negative samples is kept higher, because the LC-MS map is very sparse and most of the spaces hold no feature.

\subsection*{Deep Learning Model Design}
\begin{figure}[h!]
\centering
\includegraphics[scale = .45]{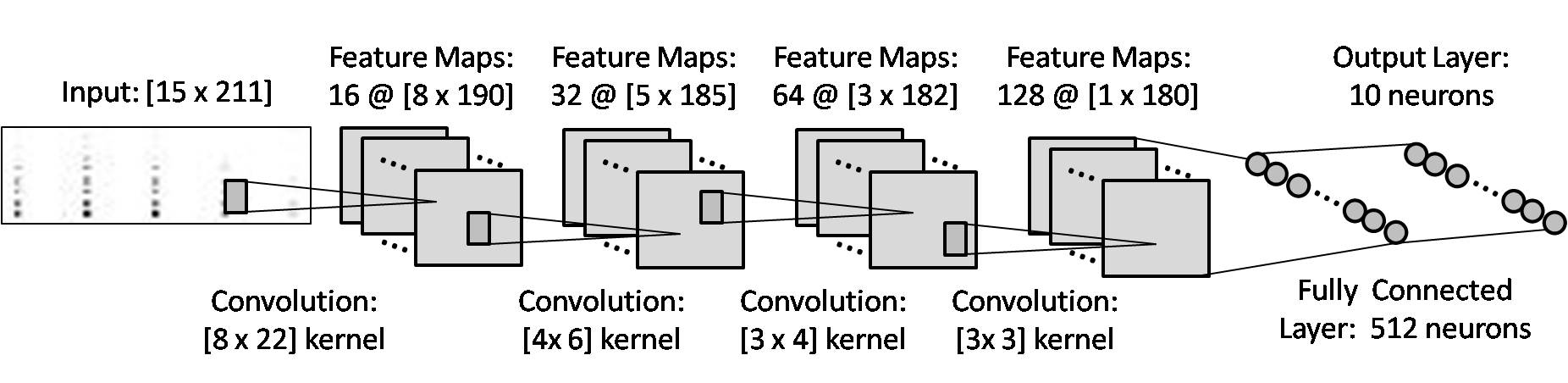}
\caption{Architecture of our proposed Convolutional Neural Network}\label{cnn}
\end{figure}

The architecture of our convolution neural network is shown in Figure~\ref{cnn}.
In order to detect the sharp boundary and location of the peptide features, we want the CNN to have the property `equivariant to translation' (ensured by CNN filters) to generalize edge, texture, shape detection in different locations, but not `invariant to translation' (ensured by pooling layers) that causes the \emph{precise location} of the detected features to matter \emph{less}. Therefore we avoid using pooling layer. We use Google developed Tensorflow library to implement our model. 
We apply stochastic optimization using Tensorflow provided `AdamOptimizer' with learning rate of $10^3$~\cite{kingma2014adam}. We use the rectified linear unit (ReLU) as activation function of the neurons and sparse softmax cross entropy as error function at the output layer\footnote{https://www.tensorflow.org/api\_docs/python/tf/nn/sparse\_softmax\_cross\_entropy\_with\_logits}. We add dropout layer after final convolution layer and fully connected layer with a value of 0.50, which 
increases the validation accuracy by  1.5\%. Minibatch size is considered 128 to ensure enough weight update in each epoch. We check the accuracy on validation set after training on each 10 minibatches. We perform data shuffling after each epoch which helps to achieve faster convergence. Our CNN converges within 30 epochs. 

\subsection{Step 2: Scan the Test LC-MS Map}
The second step starts the testing phase where the CNN trained in step 1, is given a LC-MS map of some sample. As shown in Figure~\ref{alg}, it scans the whole map in a sliding window fashion, pixel by pixel, in column major order from bottom to top. At each coordinate $(x,y)$ of the map, a window or block of dimension $[M \times N]$ starting at $(x,y)$ is fetched and fed as input image to the CNN. The CNN produces output $z=$ 0 to 9 indicating whether it sees any feature starting at that coordinate or not. We keep nine hash tables for recording the detection coordinate of $z=$ 1 to 9 classes of features during the scan. The $m/z$ values (represented by $x$ coordinate) of features are used as the key of these hash tables, and the RT ranges of the isotopes in a feature (represented by $y$ coordinate) are inserted as values under these keys. For example, in Figure~\ref{alg}, in Step 2, a feature is detected which starts at $X$ $m/z$, and first isotope has RT range from $t_1$ to $t_2$.  Interested readers are requested to have a look at the details of scanning procedure in Appendix A2. 

\subsection{Step 3: Produce a List of Detected Peptide Features}\label{step3}
In this step, we process the hash tables (resulting from Step 2) using some heuristics designed based on common peptide feature properties to produce a complete list of peptide features showing the $m/z$, and RT range of each isotopes and intensity of the feature as shown in Figure~\ref{alg}, in Step 3. For simplicity, we skip explaining the detailed procedure of processing the hash tables and include that later in Appendix A3. 
\\

\begin{figure}[h!]
\centering
\includegraphics[scale = .39]{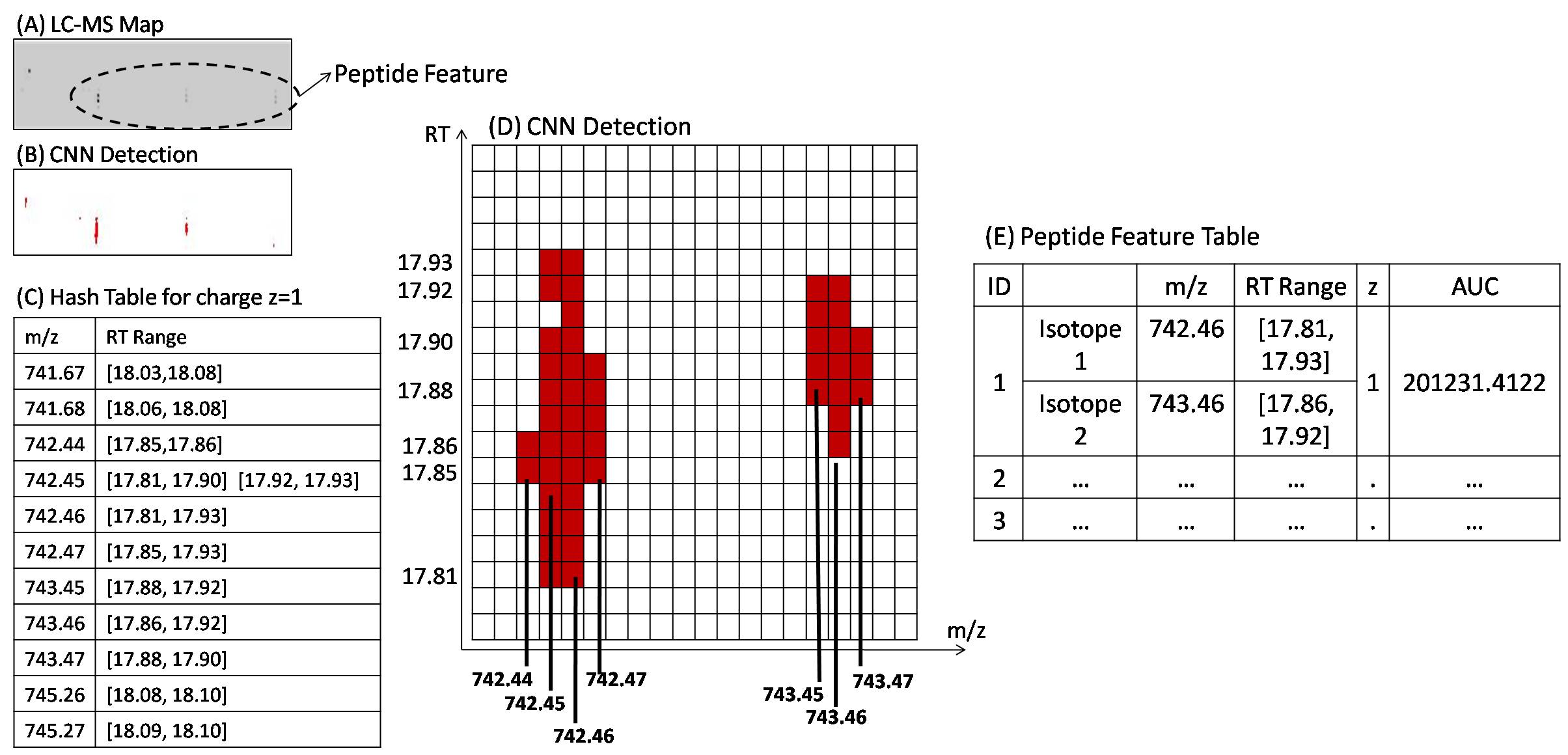}
\caption{Visualization of Step 2 and Step 3: (A) Sample LC-MS map, (B) CNN detection in Step 2 by scanning over sample LC-MS map, (C) Corresponding records in hash table after the scanning, (D) RT vs $m/z$ plot for the detected multi isotope pattern, (E) In Step 3, the pattern is listed as a peptide feature in the feature table}\label{step2n3}
\end{figure}

To clarify the intuition of test phase (Step 2 and Step 3) we show a counter example in Figure~\ref{step2n3} where a small region of LC-MS map holding a feature with charge $z=1$ is shown (in (A)) and the result of CNN detection after scanning this region (in (B)) and corresponding records in hash table (in (C)) are shown as well. Records presenting the feature is further shown using a RT vs $m/z$ plot (in (D)). After applying Step 3, the peptide feature is listed as shown in feature table (in (E)). 

\section{Result}\label{exp}
Unlike other related research works, we do not use any kind of filter to remove noises from LC-MS map before detecting features. Because we want to see how well CNN learns to avoid noises itself and detects the peptide features. Therefore, we perform all our experiments on raw MS data. 
\\

We apply a 6-fold cross validation on six LC-MS maps. Each time we keep one map for testing and the remaining are used for training. In each fold we run the training three times and choose the model giving the best validation accuracy for the 10 category classification problem. 
The average training and validation accuracy of the six folds is 94.44\% and 96.08\% respectively. The class accuracy is presented in Table~\ref{valacc}. 
\begin{table}[ht]
\centering
{\def\arraystretch{1.2}\tabcolsep=5pt
\begin{tabular}{| c | c | c|} \hline			
  charge $z$ & Training Accuracy (\%)  & Validation Accuracy (\%)\\ \hline
0&	99.87&	99.82 \\ \hline
1&	86.37&	81.80  \\ \hline
2&	93.76&	90.73 \\ \hline
3&	96.99&	94.58 \\ \hline
4&	97.12&	94.01 \\ \hline
5&	56.00 &	95.27 \\ \hline
6&	47.87 &	78.02 \\ \hline
7&	53.11 &	87.02 \\ \hline
8&	46.04&	46.07 \\ \hline
9&	45.02&	45.32 \\ \hline
\end{tabular}
}
\vspace*{2mm}
\caption{Average training and validation accuracy of six folds for each class}\label{valacc}
\end{table}

Although we use all the peptide features in training LC-MS maps for the training, however, we may not find De Novo or Database sequence for all of them. The existence of a peptide feature is supposed to be 100\% accurate if the peptide feature has a De Novo or Database sequence. The existence of other features is not wrong. However, since there is no ground truth, we do not know how accurate we are about their existence. Therefore, for the evaluation of testing phase, we arrange the features present in test LC-MS map into following two types:
\begin{itemize}
\item Type $X$: All features
\item Type $Y$: Only those features for which we can find De Novo or Database sequence. 
\end{itemize}

In testing phase, we would say a feature is detected by our proposed model DeepIso, if the feature profile reported by our method matches with the feature profile provided by PEAKS. To be specific, we compare following three points to decide whether a feature is detected.   
\begin{enumerate}
\item If the charge $z$ of a feature is matched. 
\item The $m/z$ value of the starting isotope in a feature reported by DeepIso matches with that of PEAKS result within tolerance level of 0.05 $m/z$ or Da. If the first one matches, then the remaining isotopes' $m/z$ match as well, since they are equidistant. 
\item The RT range of a feature reported by our method overlaps with that of PEAKS result and The RT value giving the highest peak intensity of a feature matches with a tolerance level of 0.5 minute.
\end{enumerate}

\subsection*{Sensitivity and Specificity of DeepIso on Test LC-MS Map}
Just like other existing works in the literature, we evaluate the True Positive rate of our model on test LC-MS map using the metric sensitivity or recall. We denote the test LC-MS map in six folds as `A', `B', `C', `D', `E', and `F' respectively. We report the sensitivity for each fold and the average sensitivity as well for data type $X$ in Table~\ref{test_acc_x}, and for data type $Y$, in Table~\ref{test_acc_y}. 

\begin{table}[ht]
\centering
{\def\arraystretch{1.2}\tabcolsep=5pt
\begin{tabular}{ |c | c | c | c | c | c | c | c |} \hline			
  charge $z$ & A & B & C & D & E & F & Average\\ \hline
  1 & 85.65 & 81.56 & 87.35 & 85.07 &	 83.68 & 85.01 &84.72\\ \hline
  2 & 90.84 & 88.68 & 88.85 & 90.80 & 89.17 & 88.10  & 89.41\\ \hline
  3 & 90.09 & 88.17 & 86.29 & 88.73 & 87.34 & 85.81 & 87.74\\ \hline
  4 & 88.53 & 86.66 & 84.90 & 89.66 & 86.80 & 83.70 & 86.71\\ \hline
  5 & 91.04 & 91.74 & 89.79 & 90.16 & 94.40 & 85.51 & 90.44\\ \hline
  6 & 71.79 & 78.49 & 73.08 & 82.99 & 78.71 & 64.89 & 74.99\\ \hline
  7 & 70.97 & 76.00 & 74.29 & 90.00 & 76.67 & 75.00 & 77.15\\ \hline
  8 & 25.81 & 28.57 & 42.86 & 37.08 & 15.38 & 16.67 & 27.73\\ \hline
  9 & 31.58 & 50.00 & 25.00 & 2.56 &	16.67 &	0 & 20.97\\ \hline
over all charges & 88.89 & 87.09 & 87.38  & 88.36 & 87.55 & 86.38 & \textbf{87.61}\\ \hline
\end{tabular}
}
\vspace*{2mm}
\caption{Sensitivity (\%) on test LC-MS map in six folds for type $X$ data}\label{test_acc_x}
\end{table}
\begin{table}[ht]
\centering
{\def\arraystretch{1.2}\tabcolsep=5pt
\begin{tabular}{ |c | c | c | c | c | c | c | c|} \hline			
  charge $z$ & A & B & C & D & E & F & Average\\ \hline
1&	No data&   50&		100&		75.00&	100&		75 & 80.00 \\ \hline
2&	95.37 &	95.00&	93.21&	94.68&	94.58&	93.80 & 94.44\\ \hline
3&	92.13 &	94.53&	88.24&	92.59&	92.68&	92.47 & 92.11\\ \hline
4&	91.11 &	92.31&	84.48&	91.14&	91.88&	84.73 & 89.28\\ \hline
5&	95.74 &	100&		91.67&	92.31&	95.96&	84.43 & 93.35\\ \hline
6&	90.48 &	100&		50&		90.57&	95.74&	87.50 & 85.71\\ \hline
  7 & No data&   No data&   No data&  No data&  No data&   No data & No data\\ \hline
  8 & No data&   No data&   No data&  No data&  No data&   No data & No data\\ \hline
  9 & No data&   No data&   No data&  No data&  No data&   No data & No data\\ \hline
over all charges & 93.98 &	94.73 &	91.06 &	93.48 &	93.75 &	92.26 & \textbf{93.21}\\ \hline
\end{tabular}
}
\vspace*{2mm}
\caption{Sensitivity (\%) on test  LC-MS map in six folds for type $Y$ data}\label{test_acc_y}
\end{table}

Detection of features having higher intensity is important in the workflow of LC-MS/MS analysis~\cite{tautenhahn2008highly}. We prepare statistics of the model sensitivity under  different intensity  range (in terms of Area Under Curve (AUC)), starting from 0 to $1\times 10^{10}$ (maximum intensity in all six LC-MS maps lie within this range). The average result of 6-fold cross validation is presented in Table~\ref{stat1} for data type $X$ and data type $Y$. Our CNN performs well when intensity is higher, as expected. 
\\
\\

\begin{table}[ht]
\centering
{\def\arraystretch{1.2}\tabcolsep=5pt
\begin{tabular}{ |c | c | c | } \hline			
  Intensity Range & Type $X$ & Type $Y$\\ \hline
  $0 \leq AUC < 1 \times 10^3$                   & 73          & 77.25 \\ \hline
  $1 \times 10^3 \leq AUC < 1 \times 10^4$ & 73.61	 & 80.06 \\ \hline
  $1 \times 10^4 \leq AUC < 1 \times 10^5$ & 81.63	& 83.82  \\ \hline
  $1 \times 10^5 \leq AUC < 1 \times 10^6$ & 88.52	& 93.04 \\ \hline
  $1 \times 10^6 \leq AUC < 1 \times 10^7$ & 92.24	 & 94.03\\ \hline
  $1 \times 10^7 \leq AUC < 1 \times 10^8$ & 94.02	& 95.13 \\ \hline
  $1 \times 10^8 \leq AUC < 1 \times 10^9$ & 96.04	& 95.93 \\ \hline
   $1 \times 10^9 \leq AUC < 1 \times 10^{10}$ & 99.24	& 100.00 \\ \hline
\end{tabular}
}
\vspace*{2mm}
\caption{The average sensitivity (\%) of 6-fold cross validation under different AUC range}\label{stat1}
\end{table}

\begin{figure}[h!]
\centering 
\includegraphics[scale = .4]{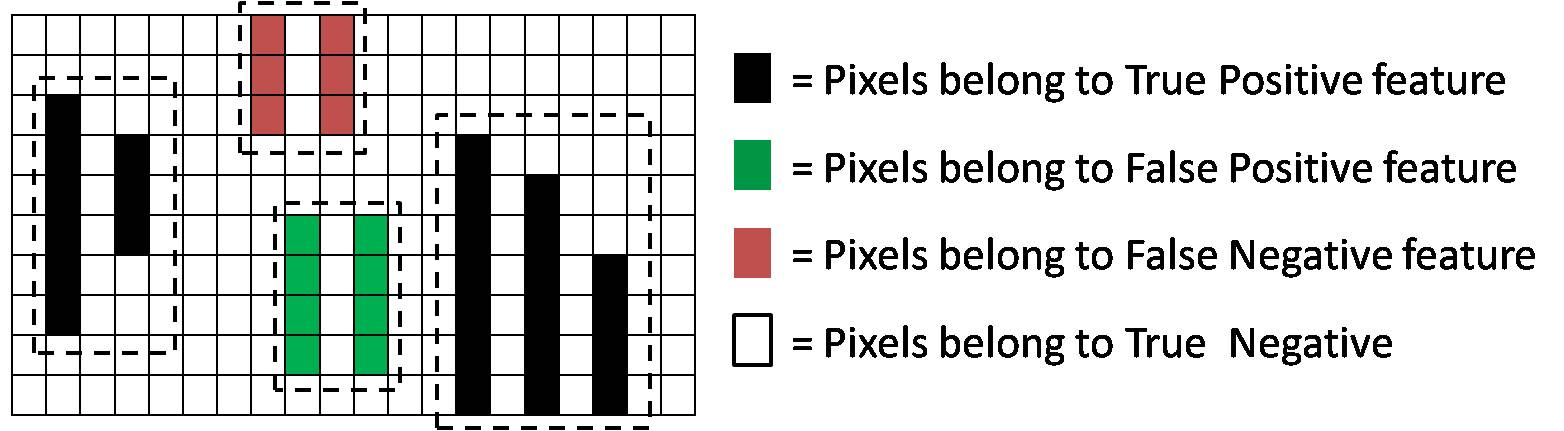}
\caption{Calculation of specificity}\label{speci}
\end{figure}

In order to evaluate the model in terms of False Positives, some of the related works in literature use the metric \emph{precision} instead of \emph{specificity} because in their methods there is no way of defining True Negative cases. However, in some works people have defined True Negative cases based on their processing strategy. For example, Conley et al.~\cite{conley2014massifquant} define True Negative cases in terms of unused centroids, and report their model specificity. Since our model finds the features in LC-MS map by scanning it using CNN, therefore the evaluation metric \emph{specificity} seems more appropriate to measure the performance of CNN. We use a counter example to explain how we define True Negatives in terms of pixels. Let us consider CNN detections after scanning over a small area of LC-MS map  having dimension [10 x 20] as shown in the Figure~\ref{speci}. Here CNN says `Yes' in the pixels shown in Black and Green. In all other pixels CNN says `No'. Now, the Black pixels belong to two True Positive features. The Green pixels represent a False Positive feature and the Red pixels represent a False Negative feature. The other White pixels represent True Negative cases. According to the figure, we can calculate the specificity in this region as follows: 
\begin{eqnarray*}
\text{specificity} & = & \frac{\text{pixels belong to True Negative (White pixels)}}{\text{pixels belong to True Negative + pixels belong to False Positive (Green pixels)}} \\
             & = & \frac{160}{168} \\
              & = &  95.24\%.
\end{eqnarray*}
In this way the average specificity of 6-fold cross validation of our model is \textbf{99.44\%}.
\subsection*{The Correlation of Peptide Intensity Between PEAKS and DeepIso}
For the statistical analysis of biological experiments the feature intensity is of interest and has to be calculated from the raw data~\cite{tautenhahn2008highly}. The technique is to first apply curve fitting over the bell shaped intensities of isotopes in a feature. Then the Area Under Curve (AUC) of all isotopes in a feature are calculated and added to get the intensity or AUC of that feature. The Pearson correlation of feature intensity calculated in our model and PEAKS is 0.78 for type $X$ data, and \textbf{0.83} for type $Y$ data (average of 6-fold cross validation).

\section{Discussion}\label{limit}
First, we would like to discuss our observations on model sensitivity as listed below.
\begin{itemize}
\item The class sensitivity varies from each other as shown in Table~\ref{test_acc_x} and Table~\ref{test_acc_y}, due to the imbalanced dataset, as reported in Table~\ref{dist}. In this table we see the distribution of features having different charges in LC-MS map `D', for both type $X$ and type $Y$.

\item The proportion of type $Y$ features having charge 1 in test LC-MS map is usually very low, as shown in Table~\ref{dist}. As a result, sensitivity of charge 1 shown in In Table~\ref{test_acc_y} does not reflect how well CNN learns to detect them. We plan to test on LC-MS map having more type $Y$ charge 1 features to evaluate the model in future.

\item Sensitivity of type $Y$ data is usually higher than type $X$, as reported in Table~\ref{test_acc_x} and Table~\ref{test_acc_y}. \emph{Sensitivity of type $Y$ data is of more significance}, because type $X$ data may contain some wrongly detected features (since PEAKS accuracy is around 95\%). However, the type $Y$ data is supposed to be 100\% accurate, since they have De Novo or Database sequence identified. Therefore, the higher sensitivity of our model for type $Y$ data indicates proper learning by the CNN model. So the average sensitivity of six folds for type $Y$ data  \textbf{93.21\%} is considered as our model sensitivity. 
\end{itemize}

\begin{table}[ht]
\centering
{\def\arraystretch{1.2}\tabcolsep=5pt
\begin{tabular}{| c | c | c|} \hline	
charge $z$ & Type $X$ & Type $Y$\\ \hline		
1&	17.46&	0.18\\ \hline
2&	39.36&	53.43\\ \hline
3&	27.78&	30.53\\ \hline
4&	7.67&	9.76\\ \hline
5&	3.43&	3.75\\ \hline
6&	2.31&	2.35\\ \hline
7&	1.10&	0\\ \hline
8&	0.61&	0\\ \hline
9&	0.27&	0\\ \hline
\end{tabular}
}
\vspace*{2mm}
\caption{Distribution (\%) of features of different charges in LC-MS map `D', for both data types}\label{dist}
\end{table}

Next, we point out some limitations in our DeepIso model and propose potential solutions as well. Based on our observations, there are two limitations with the scanning process in Step 2. First one is, the last isotope in a feature may not be recorded at all. The second one is, the ending RT values of isotopes in a feature in LC-MS map may not be recorded correctly. The reasons of these limitations are discussed in Appendix A2. Since the intensity of last isotope and around the ending trail of an isotope is usually very low, these limitations do not affect significantly in intensity (AUC) calculation. However, we propose some potential solutions to overcome these problems:
\begin{itemize}
\item To solve the first problem, we can use another CNN that is trained to scan the LC-MS map from right to left. As a result CNN would make entry to the features from the last isotope, and would be able to detect the last isotope first. 
\item  The second problem can be solved by flipping the LC-MS map along horizontal axis and let CNN scan again following usual techniques. As a result it can detect the ending RT values of the isotopes first. 
\end{itemize}
Now, we would need to combine these new detections with the old one to produce the final list of features. Another problem, the high tolerance level, might be reduced by considering 0.001 resolution along both axis $m/z$ and RT. 
\\

Finally, we discuss some potential future scopes. In our current method, we are using very basic CNN that can just output the existence of a feature along with it's charge. But it cannot directly output the region where the feature lies. Thus we have to process CNN detections applying heuristics to produce a final list of features. Therefore it will be interesting to investigate whether we can find more suitable deep neural network that not only detects but also outputs the boundary of the feature, and let us avoid heuristics completely. Some probable solutions can be R-CNN~\cite{girshick2014rich} and Mask CNN~\cite{wei2016mask}. Besides using CNN, we intend to experiment by integrating CNN and Recurrent Neural Network (RNN) into a dynamic detector~\cite{dedynamic}, which is more data-driven and can predict variable size patterns more accurately than a fix size filter. This type of models may also solve the limitations discussed. Therefore, our next concern is to design more suitable deep neural network that learns to detect the peptide features and also their boundary correctly in the LC-MS map without applying heuristics. After designing such a model we would like to perform a comparison with other existing software tools to evaluate the performance gain due to deep learning in terms of accuracy and efficiency. However, we believe that our current research reveals the capability of CNN in this domain more clearly and let us understand it's limitations as well, which would help us to design more powerful deep learning model to automate peptide feature detection in future.

\bibliographystyle{unsrtnat}
\bibliography{fatema_recomb18}

\section*{Appendix}
\subsection*{A1. Explanation of Chosen Block/Window Size}
 The hight of the block is chosen to be 15 pixels since that seems enough to discover the bell shaped intensity  of the isotopes. On the other hand, the width of block is considered 211 pixels = 2.11 m/z, because this is sufficient to detect the equidistant property of all the charges. The isotopes in features with $z=1$ are 1.0 m/z (100 pixels)  apart from each other. Usually peptide features have more than two isotopes. To look over three consecutive isotopes of a feature having charge $z=1$, window width of 211 pixels is enough. In all other charges, the isotopes are closer to each other. Therefore, this block size let the CNN look over sufficient area of peptide features to take decision about it's existence and charge. 

\subsection*{A2. Supplementary material for Step 2}
The model trained in step 1 is used to scan the given test LC-MS map pixel by pixel, in a column major order, from left to right, bottom to top. During scanning, the CNN produces output $z$ having value 0 to 9, indicating absence ($z=0$) or presence of feature ($z=$ 1 to 9). 

\begin{figure}[!tbp]
  \centering
  \begin{minipage}[b]{0.3\textwidth}
    \includegraphics[width=\textwidth]{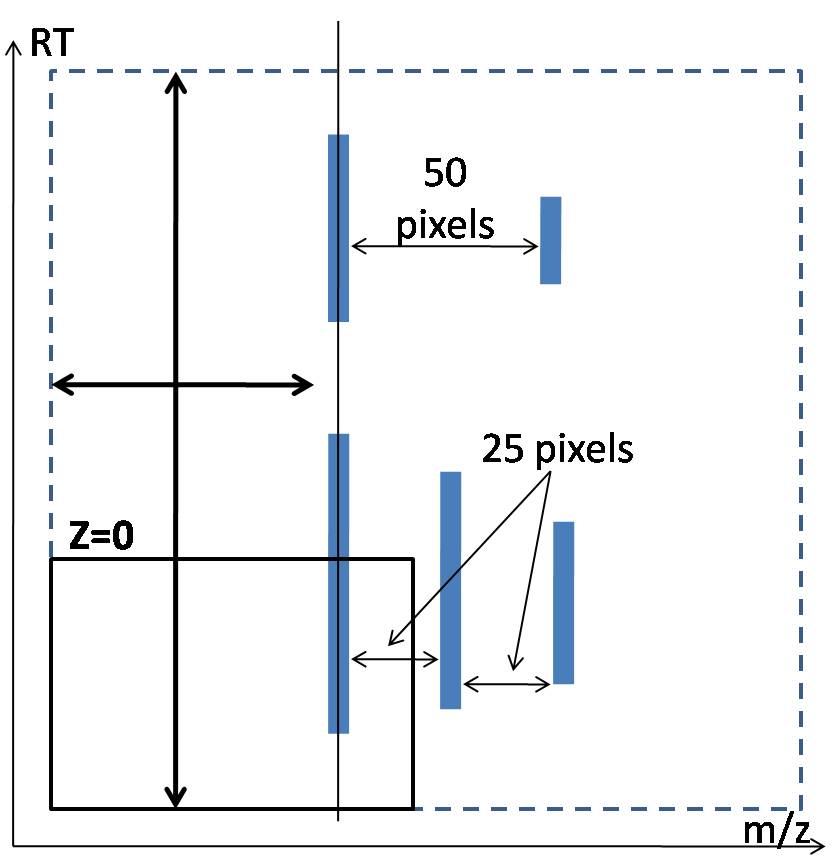}
    \caption{No feature detected}\label{sc1}
  \end{minipage}
  \hfill
  \begin{minipage}[b]{0.5\textwidth}
    \includegraphics[width=\textwidth]{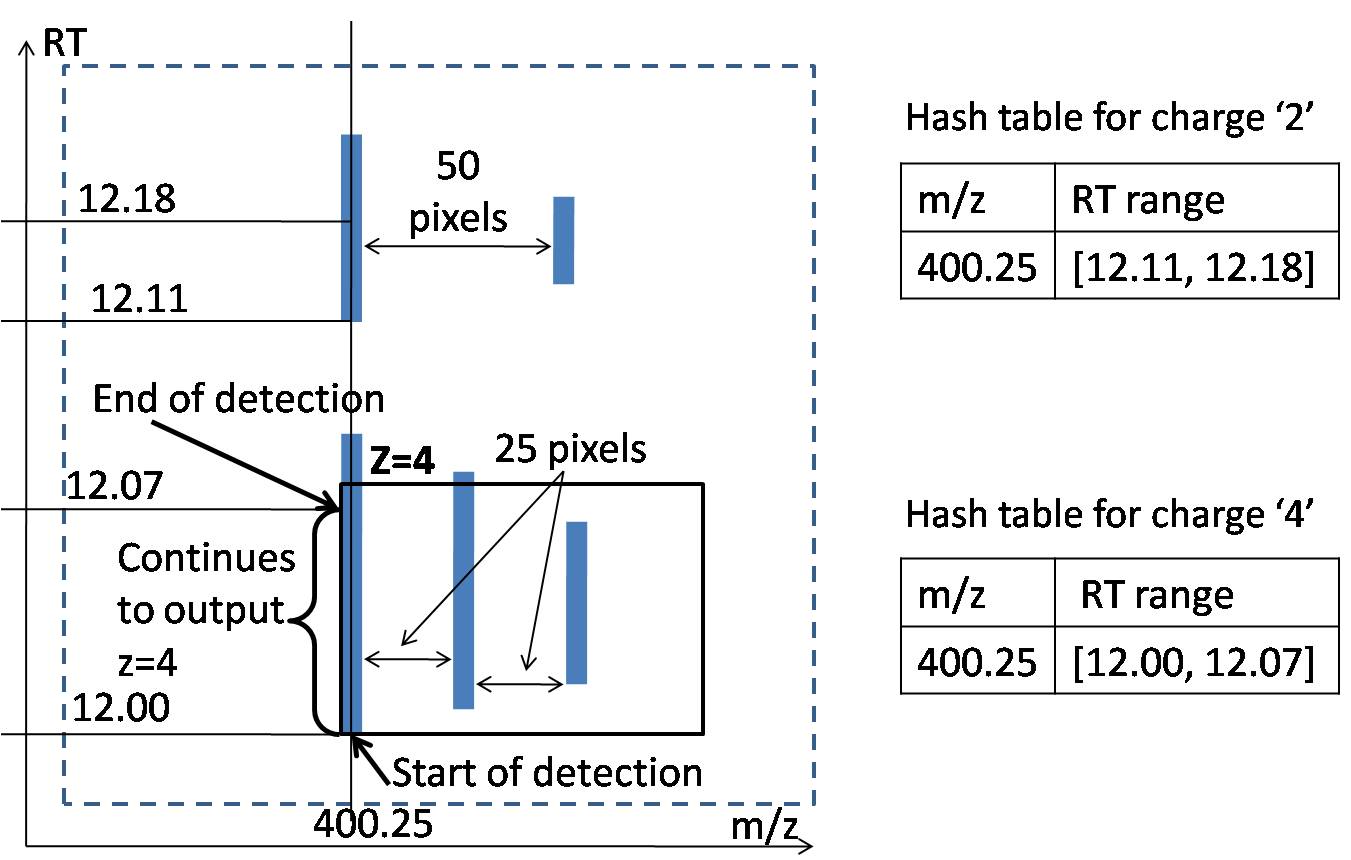}
    \caption{Detection of feature starts}\label{sc2}
  \end{minipage}
\end{figure}

\begin{figure}[!tbp]
  \centering
  \begin{minipage}[b]{0.4\textwidth}
    \includegraphics[scale=.35]{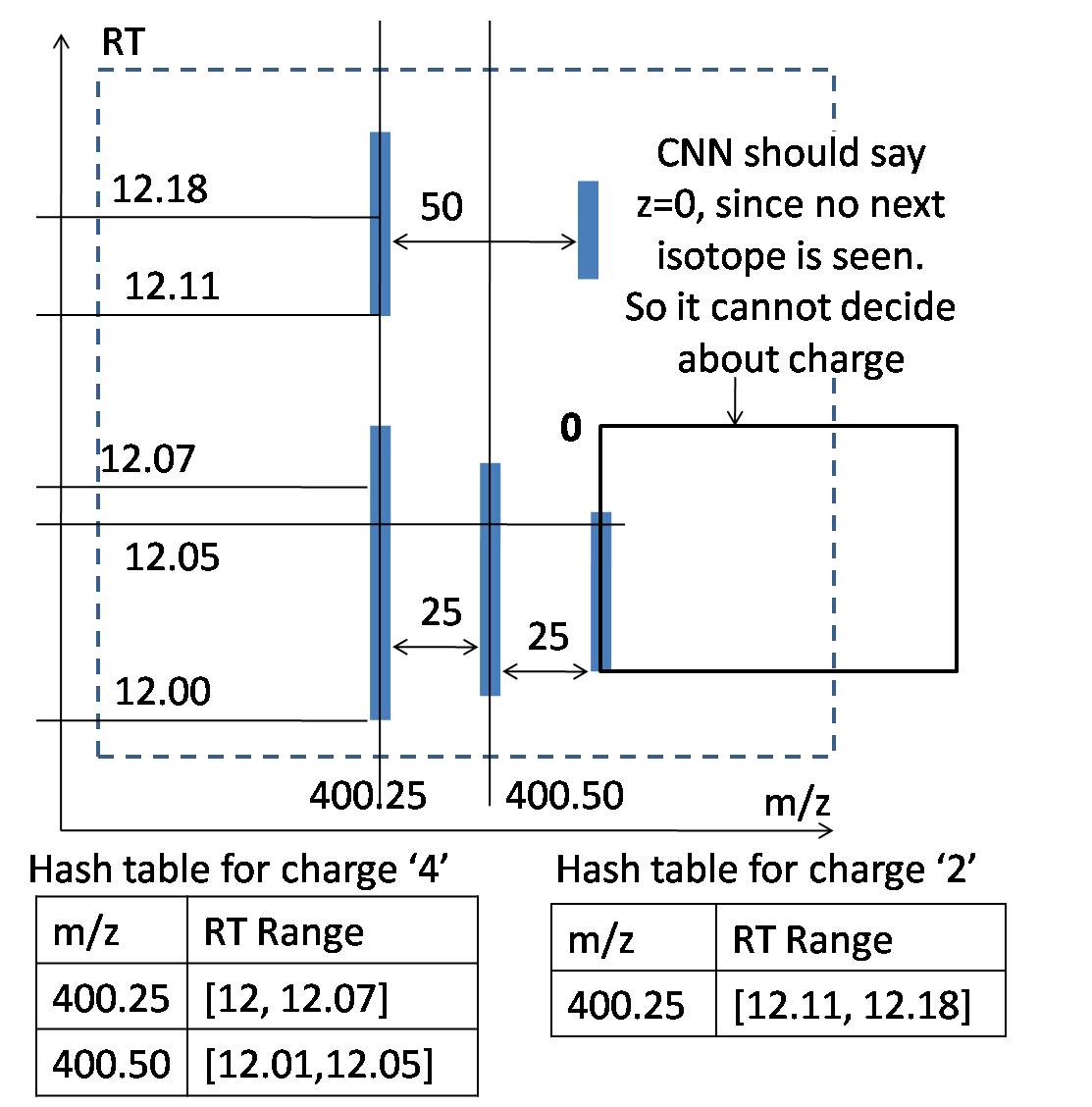}
    \caption{Two features are recorded in the hash tables}\label{sc3}
  \end{minipage}
  \hfill
  \begin{minipage}[b]{0.3\textwidth}
    \includegraphics[scale=.4]{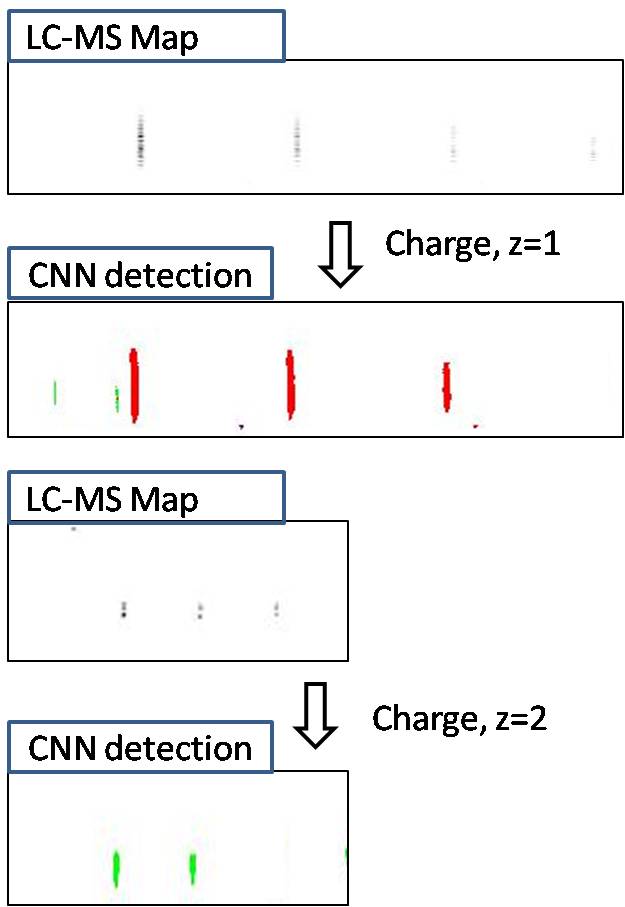}
    \caption{Feature detection by CNN}\label{detections}
  \end{minipage}
\end{figure}

To have a clarification about scanning procedure, please refer to Figure~\ref{sc1} where a small region of demo LC-MS map containing two features (charge 2 and 4) is shown. The CNN outputs $z=0$ in the area shown by arrow sign, since there is no feature. Then in Figure~\ref{sc2}, we see the position where CNN starts detecting the feature with $z=4$. We keep nine hash tables for recording the coordinate of CNN detection. So we record the starting point $(400.25, 12.00)$ in the hash table for $z=4$. Note that, we use $m/z$ value of 400.25 as key, and the RT value of 12.00 as a value under that key. The scanning outputs $z=4$ for the successive scans along that edge of the feature shown by curly brace. Therefore they indicate the continuation of same feature. So we do not insert the successive values, instead we keep track of the recent RT values and wait until CNN says $z=0$. The $(400.25,12.07)$ is the last point when CNN says $z=4$, because after that position the bell shape property and/or equidistant property of isotopes is lost. So RT $=12.07$ is inserted as the ending RT value of this feature. As a result, \emph{the ending RT values of the isotopes in a feature is not recorded correctly}. Now, the CNN continues to scan and another feature with $z=1$ is detected and it's $m/z$ and RT range $(400.25, 12.11-12.18)$ is inserted into the corresponding hash table as before. This process continues and the entries in the hash tables after the scanning over this area is done, is shown in Figure~\ref{sc3}. Please note that, \emph{the last isotope in a feature may not be detected} as well, because when the scanning window is going along the edge of last isotope, as shown in Figure~\ref{sc3}, the window does not see any other isotopes following, so it cannot decide about the charge $z$. To visualize the CNN detections, we use color Red, Green, Blue, Indigo, Lavender, Brown, Orange, Yellow, Maroon, respectively for charge $z=1, 2, 3, 4, 5, 6, 7, 8, 9$. Some detections are presented in Figure~\ref{detections}. 



\subsection*{A3. Supplementary Material for Step 3}
\begin{figure}[h!]
\centering
    \includegraphics[scale = .4]{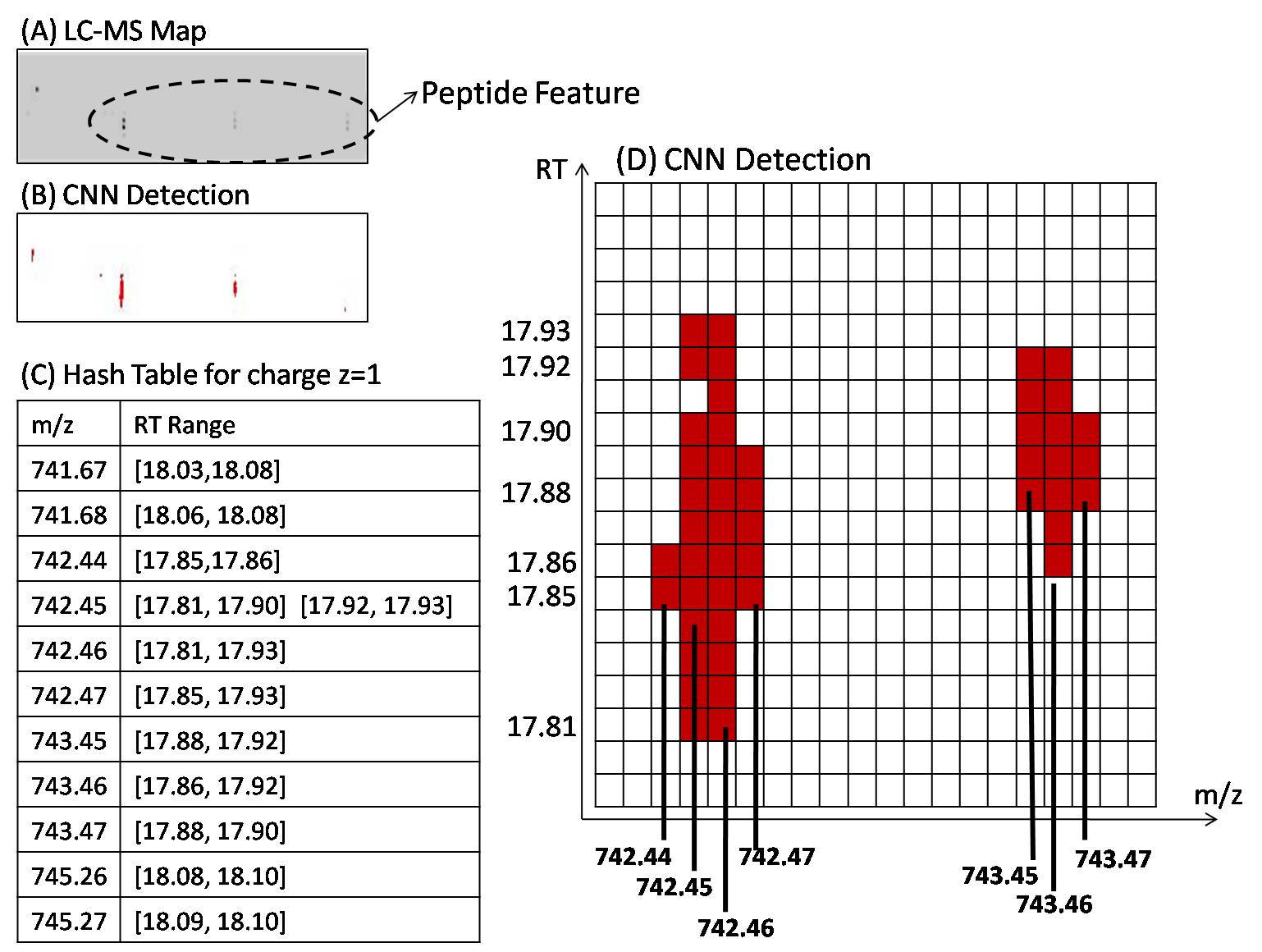}
    \caption{CNN detections recorded in hash table}\label{s31a}
\end{figure}

After CNN scans test LC-MS map, the detections recorded in hash tables are processed using some heuristics based on common feature properties, to produce a complete list of peptides showing the $m/z$, and RT range of each isotopes and intensity of the feature. Please refer to Figure~\ref{s31a} where a small region of LC-MS map holding a feature with charge $z=1$ is shown (in (A)) and the result of CNN detection after scanning this region (in (B)) and corresponding records in hash table (in (C)) are shown as well. Records presenting the feature is further shown using a RT vs $m/z$ plot (in (D)).

Following three steps are performed while processing the hash tables:
\begin{enumerate}
\item Merging the RT extents: Noise during data record causes some break in a feature as shown in Figure~\ref{brk}. 
\begin{figure}[h!]
\centering
	\includegraphics[scale = .4]{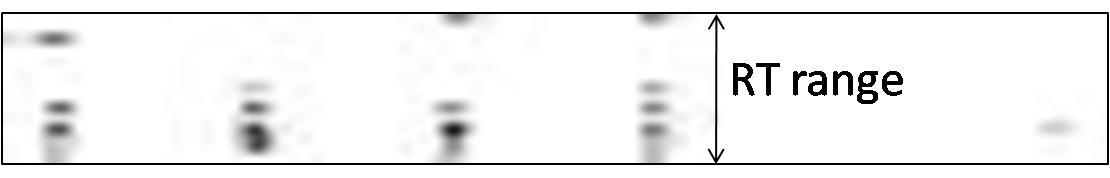}
	\caption{Break within a peptide feature}\label{brk}
\end{figure} 
Because of this the CNN detection may produce traces with small gaps as shown by arrow sign in Figure~\ref{s32}. We merge such small gaps if the gap $<= 5$ pixels, that is $0.05$ minutes. This value is chosen by experiment.

\begin{figure}[h!]
\centering
	\includegraphics[scale = .4]{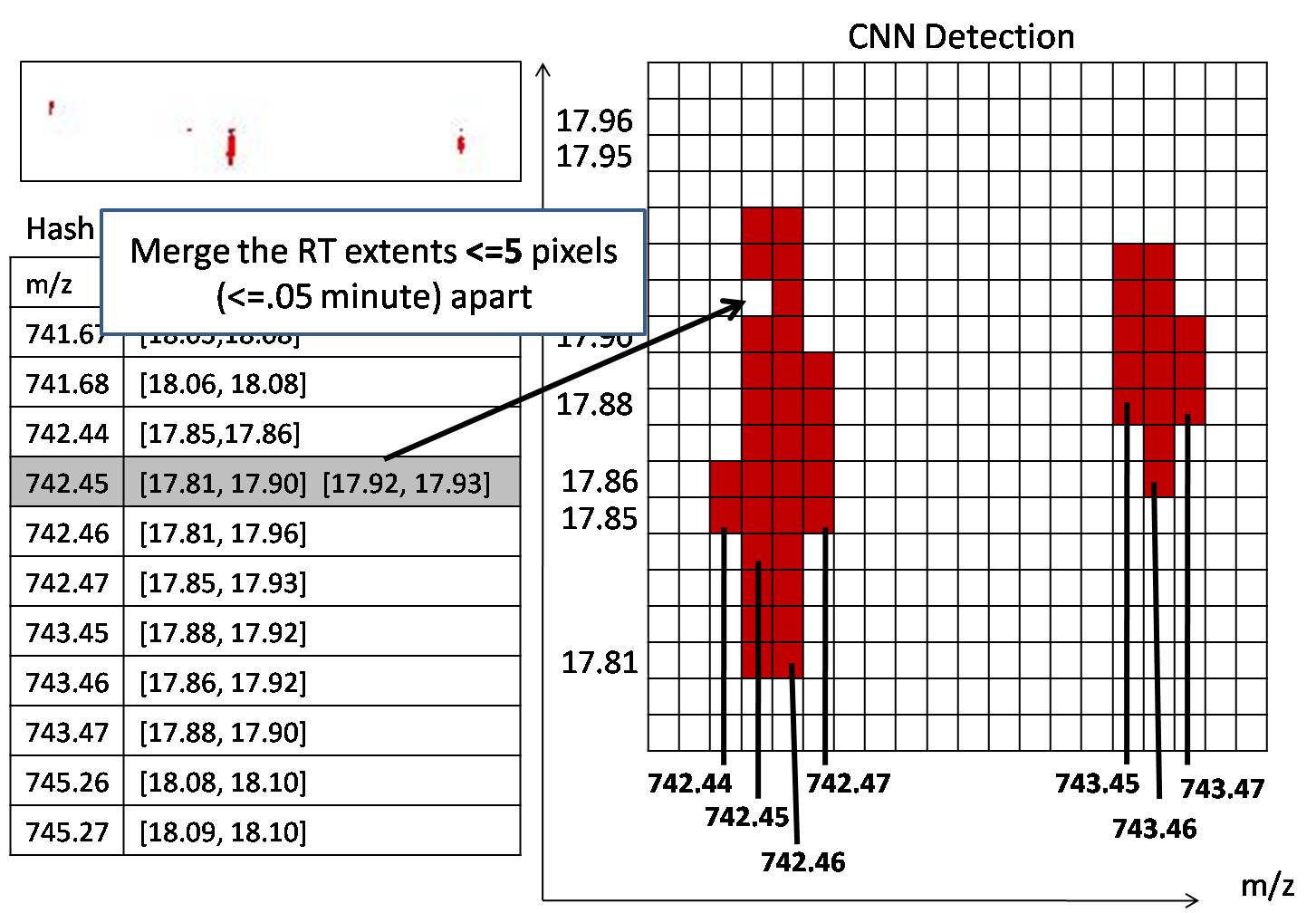}
	\caption{Merging of RT Extents}\label{s32}
\end{figure} 

\begin{figure}[h!]
\centering
	\includegraphics[scale = .4]{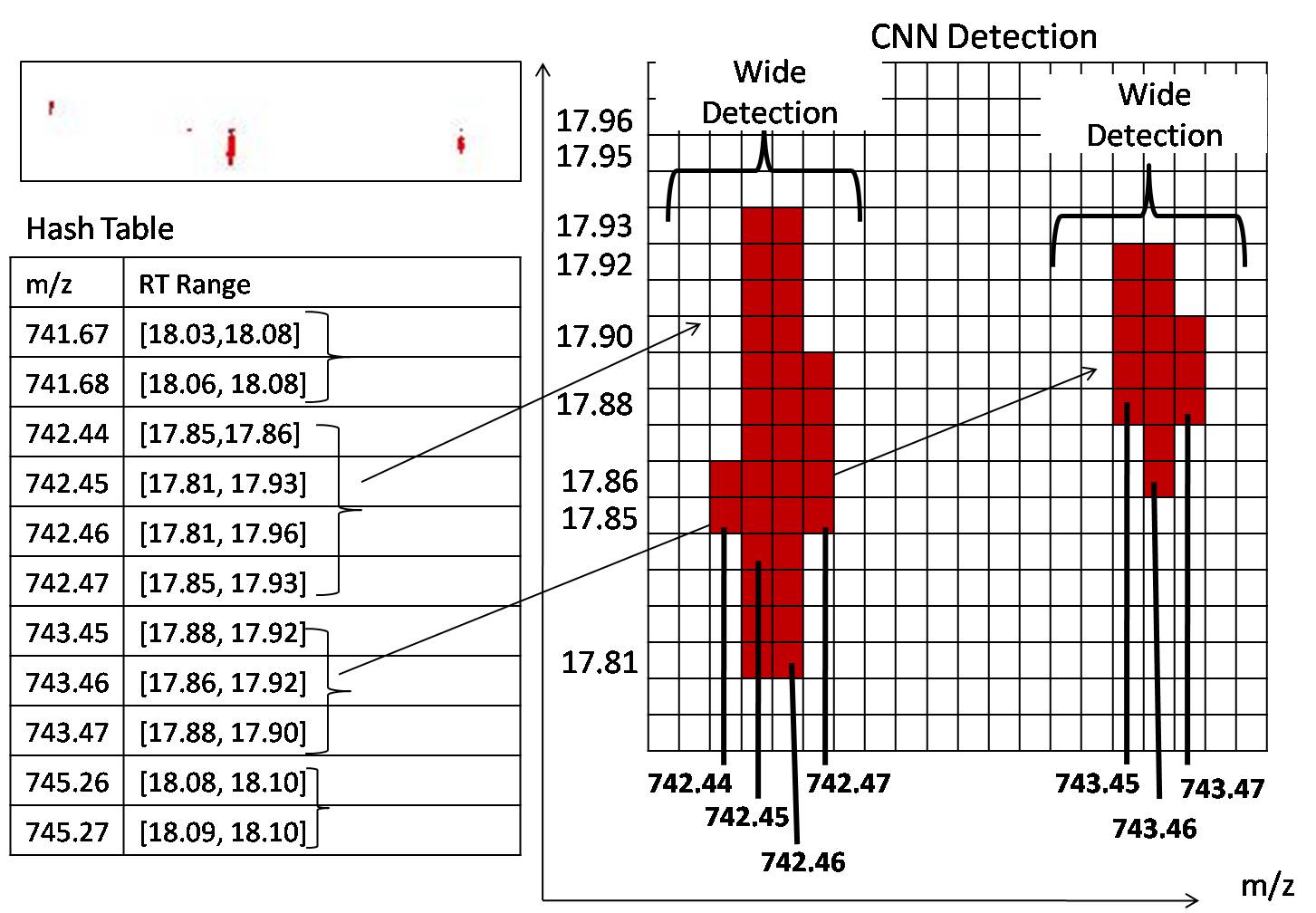}
	\caption{Combine adjacent traces who are overlapped along RT axis}\label{s33}
\end{figure}

\item Select the center m/z for a wider isotope: Although PEAKS reports just a single $m/z$ value for an isotope, but each isotope has width. That is, each isotope span over multiple pixels along $m/z$ axis. Therefore the CNN also produces wide detection as visible in Figure~\ref{s33}. However, we have to pick one m/z value for each isotope. For a set of adjacent traces representing one isotope, we calculate the intensity in terms of Area Under Curve (AUC)  for each of them, and select the one that gives highest AUC as shown in Figure~\ref{s34}.   

\begin{figure}[h!]
\centering
	\includegraphics[scale = .4]{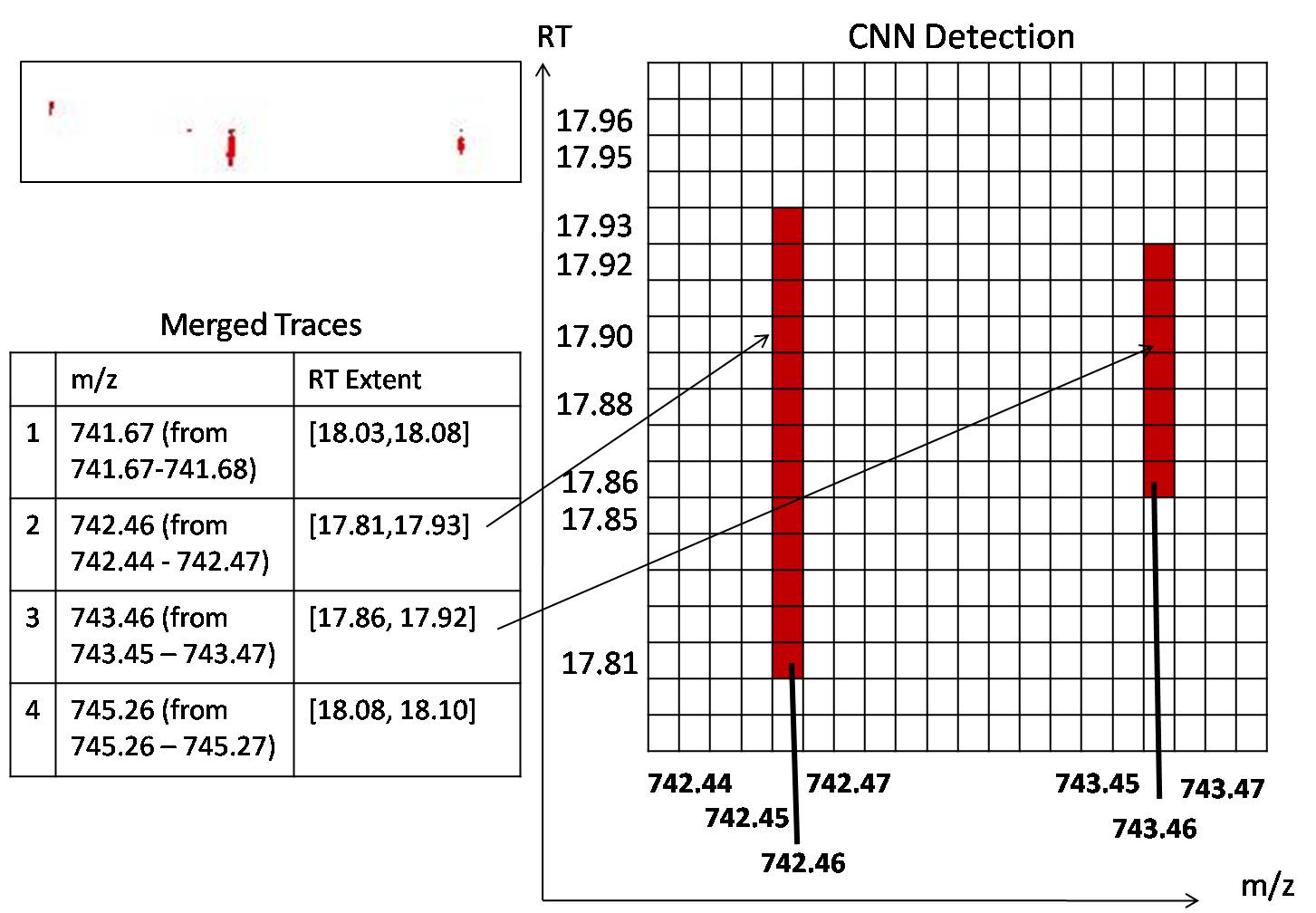}
	\caption{Selection of single m/z value for each isotope}\label{s34}
\end{figure} 

\item Combine potential isotopes into one feature: In this step we apply some heuristics as explained below.
\begin{itemize}
\item We focus on the equidistant isotope property and usual shape of features as shown in Figure~\ref{shape}. 
\begin{figure}[h!]
\centering
\includegraphics[scale = .5]{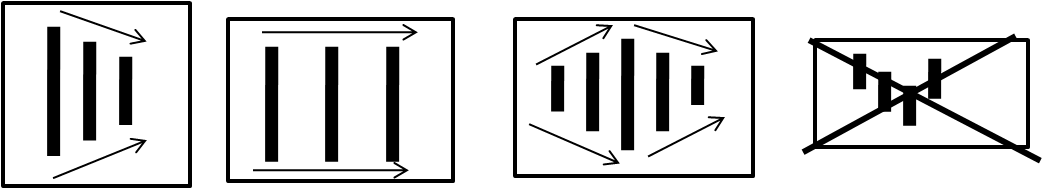}
\caption{The left most three shapes represent peptide feature but the last shape is probably noise, therefore ignored in our method}\label{shape}
\end{figure} 
\begin{figure}
  \centering
  \begin{minipage}[b]{0.4\textwidth}
\includegraphics[width=.8\textwidth]{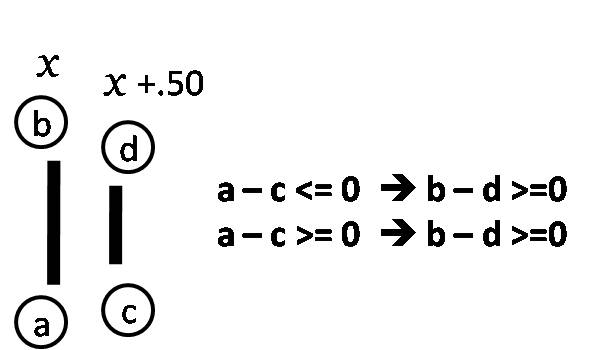}
\caption{Condition between consecutive isotopes in a feature}\label{cond}
  \end{minipage}
  \hfill
  \begin{minipage}[b]{0.4\textwidth}
    \includegraphics[width=\textwidth]{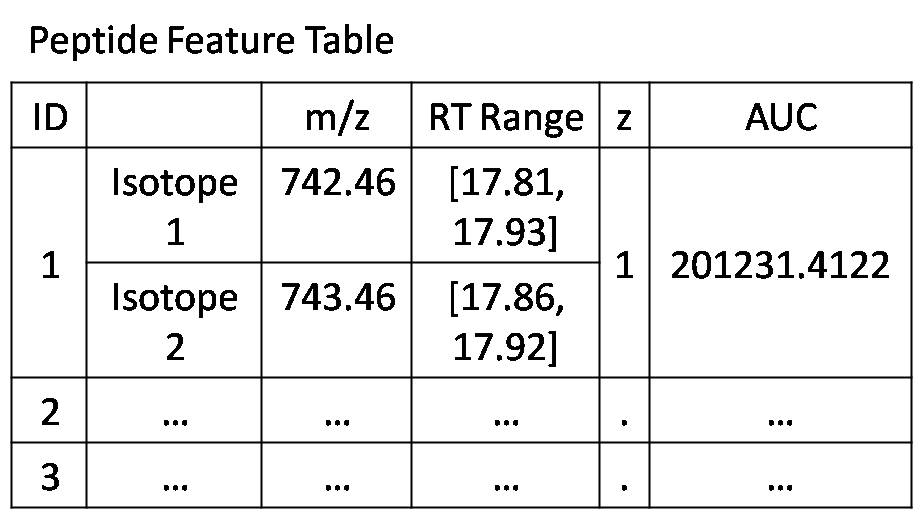}
    \caption{List of detected peptide features}\label{s35a}
  \end{minipage}
\end{figure}

Please refer to Figure~\ref{cond}, where we see a feature having charge $z=2$, and its first isotope/trace is at $x$ m/z with $(b-a)$ RT extent. It should get it's next isotope at $(x+.50)$ m/z, with overlapping RT extent $(d-c)$, satisfying the condition shown in the figure. It is found by our experiment that in 99\% cases, this relation holds between two consecutive isotopes within a same feature. We allow +-2 pixels distortion at the RT extreme points. This is also experiment based.
\item Another property is that, for two consecutive isotopes A and B in a feature: Intensity of $A >= \frac{(\text{Intensity of B})}{3}$ (or the opposite). Otherwise we consider them as belonging to two different features. 
\end{itemize}
The isotopes holding these conditions are grouped in to one feature and inserted into a final list of detected peptide features. For example, the isotopes shown in Figure~\ref{s34} are grouped in to one feature and listed as shown in Figure~\ref{s35a}. 
\end{enumerate}

\subsection*{A4. Generation of Synthetic Positive Samples for Charge $z=$ 5 to 9}
\begin{figure}[h!]
\centering
	\includegraphics[scale = .5]{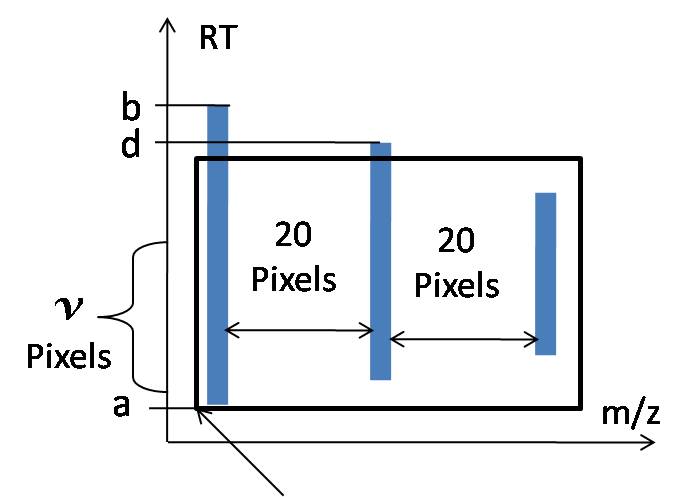}
	\caption{Synthetic positive sample generation for charge $z=$ 5 to 9}\label{syn}
\end{figure}

Please refer to Figure~\ref{syn} to understand the process of synthetic data generation. The window position shown by arrow sign is the usual position of cutting positive samples. Here we consider a feature having charge $z=5$. Based on the assumption that CNN can detect feature from this window observing the bell shaped intensity and equidistant properties of the isotopes (besides other probable hidden properties), if we slide the window in upward direction (RT axis) along the edge of first isotope of the feature, and cut images associated with those respective window positions, then the CNN should be able to detect the same feature from these additional images, as long as bell shape and equidistant properties are not lost (also apparent from the scanning process shown in Figure~\ref{sc2}). Therefore to generate synthetic positive samples, starting from the usual window position, we slide the window $v$ pixels in upward direction where $v=\min(b-a,d-a)\times \frac{2}{3}$, cut associated images from the LC-MS map, label them with the feature charge $z$, and add those to training dataset. Please note that, validation set does not contain any synthetic image.

\end{document}